\documentclass[aps,preprint,amssymb]{revtex4}

\usepackage{epsfig}

\begin{document}
\renewcommand{\thefootnote}{\fnsymbol{footnote}}
\renewcommand{\theequation}{\arabic{section}.\arabic{equation}}

\title{Physical signatures of discontinuities of the time-dependent exchange-correlation potential}

\author{Daniel Vieira,$^{a,b}$ K. Capelle$^a$ and C. A. Ullrich$^{b}$}
\email[]{E-mail: ullrichc@missouri.edu}

\affiliation{$^a$ Departamento de F\'isica e Inform\'atica, Instituto de F\'isica de S\~ao Carlos,
Universidade de S\~ao Paulo, Caixa Postal 369, S\~ao Carlos, S\~ao Paulo 13560-970, Brazil.\\
$^b$  Department of Physics and Astronomy, University of Missouri, Columbia, Missouri 65211, USA.}

\date{\today}

\begin{abstract}
The exact exchange-correlation (XC) potential in time-dependent density-functional theory (TDDFT) is known to develop
steps and discontinuities upon change of the particle number in spatially confined regions or isolated subsystems.
We demonstrate that the self-interaction corrected adiabatic local-density approximation for the XC potential
has this property, using the example of electron loss of a model quantum well system.
We then study the influence of the XC potential discontinuity in a real-time simulation of a dissociation process of an asymmetric
double quantum well system, and show that it dramatically affects the population of the resulting isolated single
quantum wells. This indicates the importance of a proper account of the discontinuities in TDDFT descriptions of
ionization, dissociation or charge transfer processes.
\end{abstract}

\maketitle

\section{Introduction}
\label{intro}

The most widely used family of exchange-correlation (XC) functionals in density-functional
theory (DFT) \cite{hk,ks,dft}, the local density approximation (LDA) and its
gradient-corrected semilocal relatives, has been largely responsible for
the enormous popularity of the theory for electronic-structure calculations.
Despite many successes, there are important properties of the exact XC
potential that are missed by the LDA and many other popular approximations.
In this paper, we shall deal with one such property, namely the discontinuity
of the XC potential upon change of particle number \cite{perdew1982}.

It is well known that the exact XC energy
functional must vary linearly as a function of the total number of
particles $N$, displaying a derivative discontinuity every time the
system passes through an integral value of $N$ \cite{perdew,perdew06}.
The delocalization error \cite{yang}, however, causes the LDA to predict a
nonlinear curvature for the change of XC energy with particle number.
This has important practical consequences, for instance leading to a description of molecular
dissociation where the resulting isolated atoms end up with unphysical fractional electron numbers.
Therefore, attempts to model molecular dissociation processes, or
any phenomenon that is associated with the transport of charges between well-separated spatial
regions or subsystems, must take the discontinuity of the  XC functional into account \cite{tozer}.

In the framework of time-dependent density-functional theory (TDDFT) \cite{rg,book}, discontinuities
of the time-dependent XC potential upon change of particle number have recently been observed
numerically \cite{lein,kummel}, and shown to be essential for a proper description
of strong-field ionization processes. In such processes featuring rapid electron escape,
the formation of steps in the XC potential was observed, which eventually
develop into fully-fledged discontinuities. These previous time-dependent
studies dealt with one-dimensional model atoms, either putting in the potential discontinuity
by hand \cite{lein} and studying its influence on single versus double ionization probabilities,
or calculating the exact XC potential, including steps and discontinuities, by inverting an exact
solution of the time-dependent Schr\"odinger equation \cite{kummel}.

In this paper we shall work along somewhat different lines: our primary interest is in
the question how these features of the time-dependent XC potential influence the
behavior of a dissociating system. Rather than carrying out a numerically
exact calculation of a simple, few-electron model system, we shall work with physically
more realistic (yet still simple) systems, namely doped semiconductor quantum wells in effective-mass approximation, and study the
electron loss of single quantum wells, and the dissociation of an asymmetric double
quantum well system. In our TDDFT simulations of these processes, we shall
compare the adiabatic LDA with and without two types of self-interaction
correction (SIC). We will first
show that the LDA plus SIC indeed leads to jumps in the XC potential under the appropriate
conditions. Then, we will investigate how these jumps affect the dissociation of the
double well system.

The paper is organized as follows: Section 2 gives some theoretical background
on the concept of self-interaction error and its correction in DFT and TDDFT,
as well as the basics of our quantum well model system. In
Sections 3 A and B we present results of time-dependent simulations of
electron loss and dissociation of single and double quantum wells,
respectively. Finally, in Section 4 we give our conclusions.

\section{Theoretical Background}
\label{theorback}
\subsection{Static and dynamic self-interaction correction}

In static spin-DFT (SDFT), the noninteracting Kohn-Sham (KS) electrons move
in a spin-dependent effective potential given by
\begin{equation}
\label{vsstatic} v_{\sigma}[n]({\bf r}) = v_{\rm ext,\sigma}({\bf r})+
v_{\rm H}[n]({\bf r}) + v_{\rm  xc,\sigma}[n_\uparrow,n_\downarrow]({\bf r}).
\end{equation}
The density is obtained  via $n({\bf
r})= \sum_{\sigma=\uparrow,\downarrow}n_\sigma({\bf r}) =
\sum_{\sigma=\uparrow,\downarrow}\sum_k^{occ}n_{k\sigma}({\bf r})$, where
the orbital densities are related to the KS orbitals by $n_{k\sigma}({\bf r})= |\varphi_{k\sigma}({\bf
r})|^2$. The first term in Eq. (\ref{vsstatic}) is the external potential, followed by the
Hartree and the XC potential.

One of the exact conditions that the XC
potential must satisfy is to be free of the self-interaction error. This means, in particular, that the Hartree
and XC potential must cancel each other in a one-electron system. This is
automatically true if, for instance, we approximate $v_{\rm  xc,\sigma}$ by the exact-exchange (EXX)
potential \cite{kummel08}. However, for LDA and semilocal functionals this
is a hard constraint to be satisfied. Therefore, various procedures for
removing the self-interaction error of an approximate XC functional have been
developed. Perdew and Zunger \cite{pzsic}
proposed the most widely known SIC for approximate XC energy functionals:
\begin{eqnarray}
E_{\rm xc}^{PZSIC}[n_\uparrow,n_\downarrow]=E_{\rm xc}^{approx}[n_\uparrow,n_\downarrow]
- \sum_{\sigma=\uparrow,\downarrow} \sum_k^{occ}
\left(E_{\rm H}[n_{k\sigma}] + E^{approx}_{\rm xc}[n_{k\sigma},0]\right).
\label{pzsic}
\end{eqnarray}
For a one-electron system it is immediately seen that the XC contributions
cancel each other, and $E_{\rm xc}^{PZSIC}$ becomes equal to the negative of
the Hartree energy, thus cancelling $E_{\rm H}$.

The XC potential associated with $E_{\rm xc}^{PZSIC}$, formally defined by
\begin{equation}
v_{\rm xc, \sigma}^{PZSIC}[n_\uparrow,n_\downarrow] ({\bf r}) =
\frac{\delta E_{\rm xc}^{PZSIC}[n_\uparrow,n_\downarrow]}{\delta n_\sigma ({\bf r})} ,
\end{equation}
is not straightforward to obtain by direct variation with respect to the density, since
$E_{\rm xc}^{PZSIC}$ is an explicit functional
of the orbital densities but only an implicit functional of $n_\sigma$ \cite{kummel08}.
Instead, to obtain a local, state-independent PZSIC XC potential
one must apply
the so-called optimized-effective potential (OEP) \cite{oep} methodology or one
of its simplifications such as the Krieger-Li-Iafrate (KLI)
approach \cite{kli}. The latter is defined as follows:
\begin{equation}
\label{kli}
v_{\rm xc, \sigma}^{KLI}({\bf r}) = \frac{1}{2 n_\sigma ({\bf r})} \sum_{k=1}^{occ} |\varphi_{k\sigma}({\bf r})|^2
\left\{ u_{{\rm xc}, k\sigma} ({\bf r}) + \left[\bar{v}_{{\rm xc}, k\sigma}^{KLI} - \bar{u}_{{\rm xc}, k\sigma} \right] \right\} + \textrm{c.c.} ,
\end{equation}
where
\begin{equation}
\label{uxc}
u_{{\rm xc},k \sigma} ({\bf r}) = \frac{1}{\varphi_{k\sigma}^{*}({\bf r})}\frac{\delta E_{\rm xc}^{PZSIC}[n_\uparrow,n_\downarrow]}
{\delta \varphi_{k\sigma}({\bf r})}
\end{equation}
and the bars denote orbital averages, e.g. $\bar{v}_{{\rm xc}, k
\sigma}^{KLI} =  \int d^3r |\varphi_{k\sigma}({\bf r})|^2v_{\rm xc,
\sigma}^{KLI}({\bf r})$. Being self-interaction free, the PZSIC
formulation also ensures the correct $-1/r$ asymptotic behavior for
the XC potential, contrasting with the exponential decay predicted
by the LDA. Such a feature is particularly important
when dealing, e.g., with the delocalization error \cite{yang}, since in
LDA the outermost electrons are not bound strongly enough. Moreover,
it has been shown  that the OEP/KLI formalism is able to
predict the expected potential discontinuity when the particle number
crosses an integer \cite{kli}.

The PZSIC approach is by no means the only method to get rid of self-interaction errors.
A more recent version of SIC, proposed by Lundin and Eriksson \cite{lesic1,lesic2}, removes the self-interaction
directly in the effective potential, which in this approach is given by
\begin{equation}
v_{{\rm xc},k\sigma}^{LESIC}[n_\sigma,n_{\bar{\sigma}}]({\bf r})
=- v_{\rm H}[n_{k\sigma}]({\bf r})
 +  v^{approx}_{\rm xc}[n_\sigma-n_{k\sigma},n_{\bar{\sigma}}]({\bf r}),
\label{lesicV}
\end{equation}
where the XC potential acting on the $k$th orbital depends on all orbital densities except the $k$th.
Just like PZSIC, the LESIC approach produces a state-dependent XC potential, which also requires the implementation of
the OEP/KLI procedure in order to obtain a local multiplicative KS
potential. However, contrary to PZSIC, the LESIC XC approach is formulated directly for the potential,
instead of for the energy, so that in principle we cannot use Eq. (\ref{uxc}) of the OEP/KLI formalism. Nevertheless, one can {\em assume} that
there exists an energy functional (even though we do not know its
structure) which gives rise to the LE orbital potential
(\ref{lesicV}), and adopt $u_{{\rm xc},k \sigma} ({\bf
r})=v_{{\rm xc},k\sigma}^{LESIC}[n_\sigma,n_{\bar{\sigma}}]({\bf r})$ in
Eqs. (\ref{kli}) and (\ref{uxc}). Other features such as the
asymptotic behavior and potential discontinuity are also recovered
by LESIC, but unlike PZSIC it predicts an erroneous correction even for the
hypothetic exact functional \cite{perdew05}.

Let us now consider dynamical situations. In TDDFT, the time-dependent analog of
 Eq. (\ref{vsstatic}) is
\begin{equation}
\label{vstd} v_{\sigma} [n]({\bf r},t) = v_{\rm ext,\sigma}({\bf r},t)+
v_{\rm H}[n]({\bf r},t) + v_{\rm xc,\sigma}[n_\uparrow,n_\downarrow]({\bf r},t),
\end{equation}
where $n({\bf r},t)=\sum_{\sigma=\uparrow,\downarrow} \sum_k^{occ}|\varphi_{k\sigma}({\bf
r},t)|^2$ follows from the  time-dependent KS
orbitals. $v_{\rm xc}[n]({\bf r},t)$  is an even more complex object than the static XC potential, since
at any given time $t$ it must retain memory effects
from all previous times $t'\leq t$. Modeling this is a formidable
task, and many commonly used TDDFT approaches therefore make use of the
adiabatic approximation, which takes an approximate XC potential from static
DFT and evaluates it at the instantaneous time-dependent
density $n({\bf r},t)$ \cite{adiabatic}. Accordingly, our previous equations
(\ref{vsstatic})-(\ref{lesicV}), dealing with static SIC and KLI
calculations, are readily brought under the adiabatic time-dependent
umbrella by simply replacing $({\bf r}) \rightarrow ({\bf r},t)$.

Clearly, adiabatic approximations misrepresent the
history of the system, because the resulting TD functional does not have a
memory of the past. One way to build memory in the functional has been
outlined in reference \cite{tdgcm}. Here, however, we will focus on aspects
where memory effects are not expected to be crucial. Therefore, all
time-dependent calculations we present in the following will make
use of the adiabatic approach.

\subsection{Semiconductor quantum wells}
\label{semiq}

Testing density functionals in a well controlled and modeled
environment is of crucial importance for the development of
(TD)DFT. In this sense, the quasi two-dimensional electron gases which occur in doped semiconductor heterostructures
are useful since they represent an essentially one-dimensional problem: in the
effective-mass approximation,
only the direction of quantum confinement needs to be analyzed, whereas the lateral or in-plane electronic
degrees of freedom  can be treated analytically.
In particular, GaAs/Al$_x$Ga$_{1-x}$As  heterostructures
have attracted considerable attention since the effective-mass
approximation for conduction band electrons works very well in these systems \cite{proetto2,proetto,ullrich}.

We assume that such  quantum wells have been grown along the
$z$ direction and are translationally invariant in the $x-y$ plane.
Considering only the conduction band, the KS eigenfunctions can be written as
\begin{equation}
\psi_{q_\parallel, j\sigma} ({\bf r}) = \frac{1}{\sqrt{A}} e^{i {\bf q}_\parallel \cdot {\bf r}_\parallel} \varphi_{j\sigma} (z),
\end{equation}
where $A$ is an area and $j$ is a subband index, and $\bf r_{||}$ and $\bf q_{||}$ denote in-plane position
and wave vectors.
The envelope functions $\varphi_{j\sigma} (z)$ satisfy the following one-dimensional KS equation:
\begin{equation}
\left[-\frac{1}{2m^*} \frac{d^2}{dz^2} + v_{\rm ext}(z)+ v_{\rm H}[n](z) + v_{\rm xc,\sigma}[n](z) \right] \varphi_{j\sigma} (z) = \varepsilon_{j\sigma} \varphi_{j\sigma} (z)
\end{equation}
with $n(z) = \sum_{\sigma=\uparrow,\downarrow} n_\sigma(z)= \sum_{\sigma=\uparrow,\downarrow}\sum_j (\mu -
\varepsilon_{j\sigma})|\varphi_{j\sigma} (z)|^2/4 \pi$, and $\mu$ being the conduction band
Fermi level ($\hbar=m=1$ from this point on, so that the effective mass
$m^*$ is a pure number; further, the bare electric charge $e$ is replaced by an effective charge $e^*$). The external potential
$v_{\rm ext}(z)$ is prescribed by the design of the quantum well. To
describe the dynamics we then propagate the subband envelope functions
$\varphi_{j\sigma}(z)$ using the time-dependent KS equation
\begin{equation}
i \frac{\partial}{\partial t} \varphi_{j\sigma}(z,t) = \left[-\frac{1}{2m^*} \frac{d^2}{dz^2} + v_{\rm ext}(z,t)+
 v_{\rm H}[n](z,t) + v_{\rm xc,\sigma}[n](z,t) \right]
\varphi_{j\sigma}(z,t)
\end{equation}
with the initial condition $\varphi_{j \sigma}(z,t_0)=\varphi_{j\sigma}(z)$. We use
LDA+SIC for $v_{\rm xc,\sigma}[n](z,t)$, and the corresponding TDKLI potential
\cite{tdkli} can be written as
\begin{equation}
\label{klitd}
v_{\rm xc, \sigma}^{KLI}(z,t) = \frac{1}{2 n_\sigma (z,t)} \sum_{j=1}^{occ} \frac{(\mu - \varepsilon_{j\sigma})}{4\pi}|\varphi_{j\sigma}(z,t)|^2
\left\{ u_{{\rm xc}, j\sigma} (z,t) +
\left[\bar{v}_{{\rm xc}, j\sigma}^{KLI}(t) - \bar{u}_{{\rm xc}, j\sigma} (t)\right] \right\} + \textrm{c.c.}
\end{equation}
with, e.g., $\bar{v}_{{\rm xc}, j \sigma}^{KLI}(t) =  \int d^3r
|\varphi_{j\sigma}(z,t)|^2v_{\rm xc, \sigma}^{KLI}(z,t)$. As we will demonstrate below, the
 TDKLI potential (\ref{klitd}) retains desirable
features of the exact SIC, such as the potential discontinuity
each time a new subband $j$ is filled and  $-1/z$ asymptotic
behavior. In the following, the KS orbitals are propagated on a
real-space grid in real time with a Crank-Nicholson algorithm and a
time step of 0.05 a.u.

\section{Results and Discussion}

\subsection{Barrier suppression of a single well}

Mundt and K\"ummel \cite{kummel} showed for a one-dimensional lithium model
atom that the time-dependent EXX potential displays humps at its edges
as the system loses electrons in an
ionization process. In the limit of complete ionization, these humps
become more and more steplike, reaching an asymptotic constant value
corresponding to the intrinsic derivative discontinuity of the XC
potential as the particle number passes through an integer value.
Here we study an analogous situation, letting electrons escape from a semiconductor quantum well.

\begin{figure}
  \begin{center}
\includegraphics{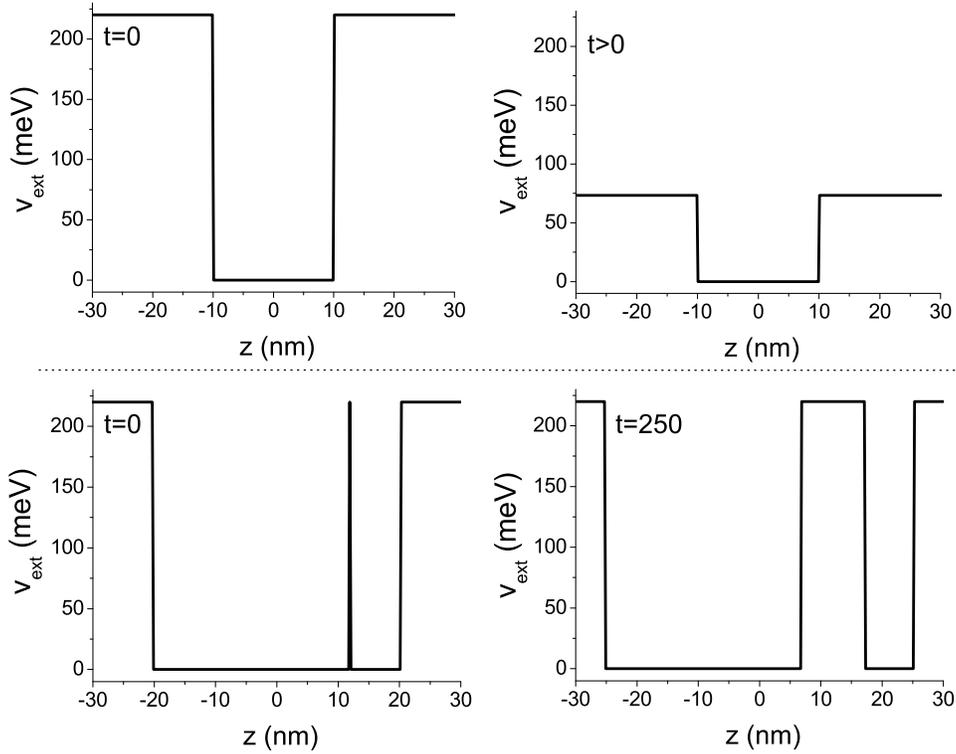}
 \caption{Upper panel: quantum well geometry for studying the effects
of barrier suppression. At time $t=0$, the well depth is abruptly reduced to $1/3$ of the initial value.
Lower panel: quantum well geometry for studying
double-well dissociation. The barrier width grows steadily from 0.2 nm at $t=0$ to 10 nm at $t=250$ a.u. \label{fig1}}
  \end{center}
\end{figure}

We consider a $20$nm square GaAs/Al$_{0.3}$Ga$_{0.7}$As quantum well
with conduction band effective mass $m^*=0.067$ and charge
$e^*=e/\sqrt{13}$, and with an electronic density such that the two
lowest subbands ($j=1,2$) are occupied. Initially ($t=0$), the
difference between the conduction band edges of GaAs and
Al$_{0.3}$Ga$_{0.7}$As is taken as 220meV. Once we have established the
KS ground states $\varphi_{j\sigma}(z,t=0)$, for $t>0$ we
suppress the barrier height to one third of the initial value. In other words,
we consider a sudden switching process which reduces the quantum well depth
by 66\%, but keeps the square well shape otherwise intact (see upper panel of
Fig. \ref{fig1}). Such a process is, of course, highly idealized and would
be difficult to realize experimentally. In practice, a static
electric field could be used to suppress the quantum well barrier on one side.

Applying an absorbing boundary condition, we then let the system
evolve such that the electrons in the second subband are almost
completely ionized, and focus our attention on the behavior of the LDA
and SIC potentials. We limit ourselves in the following to the exchange-only LDA, $v_{\rm xc,\sigma}^{LDA}[n_\sigma,
n_{\bar{\sigma}}](z,t) = -\left[6/\pi\ n_\sigma(z,t)\right]^{1/3}$, applying
the TDKLI/PZSIC and TDKLI/LESIC approaches.

\begin{figure}
  \begin{center}
\includegraphics{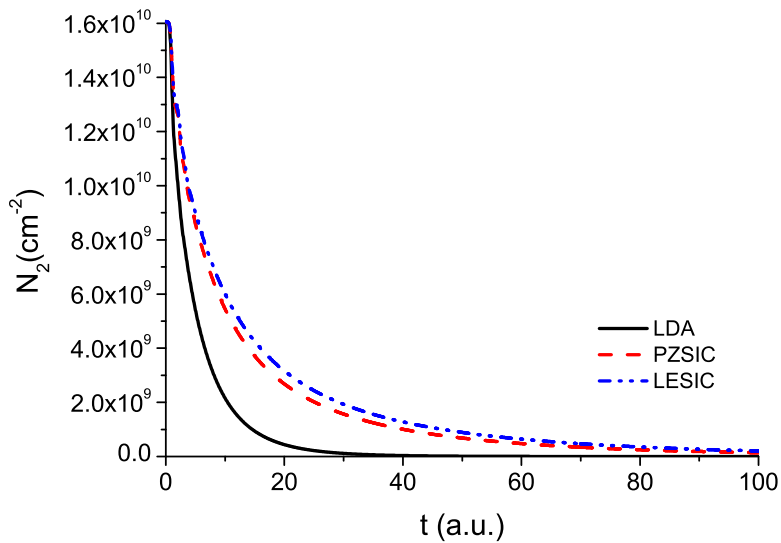}
 \caption{[Color online] Number of electrons in the second subband versus time, illustrating electron escape
following a suppression of the quantum well depth to one third of its initial value at $t=0$.
\label{fig2}}
  \end{center}
\end{figure}

Figure \ref{fig2} shows  the number of electrons per unit area in the second
subband, $N_2(t)= (2\pi)^{-1}(\mu - \varepsilon_{2\sigma}) \int |\varphi_{2\sigma}(z,t)|^2 dz$,
as a function of time. Due to the lowering of the barrier, electrons are able to escape,
move away from the quantum well and are absorbed at the boundary,
so that the norm of the envelope functions steadily decreases,
$\int |\varphi_{2\sigma}(z,t)|^2 dz<1$.
The electrons in the first subband are much more tightly bound, so that
the overall ionization of the well is almost exclusively a
consequence of emptying the second subband. As one can observe, for
pure LDA the ionization occurs much faster than for LDA+PZSIC and LDA+LESIC.
This feature is a consequence of the delocalization error \cite{yang}
and the
incorrect asymptotic behavior of the LDA: since the electrons are not
strongly enough bound to the system, they escape more easily.

\begin{figure}
\begin{center}
\includegraphics{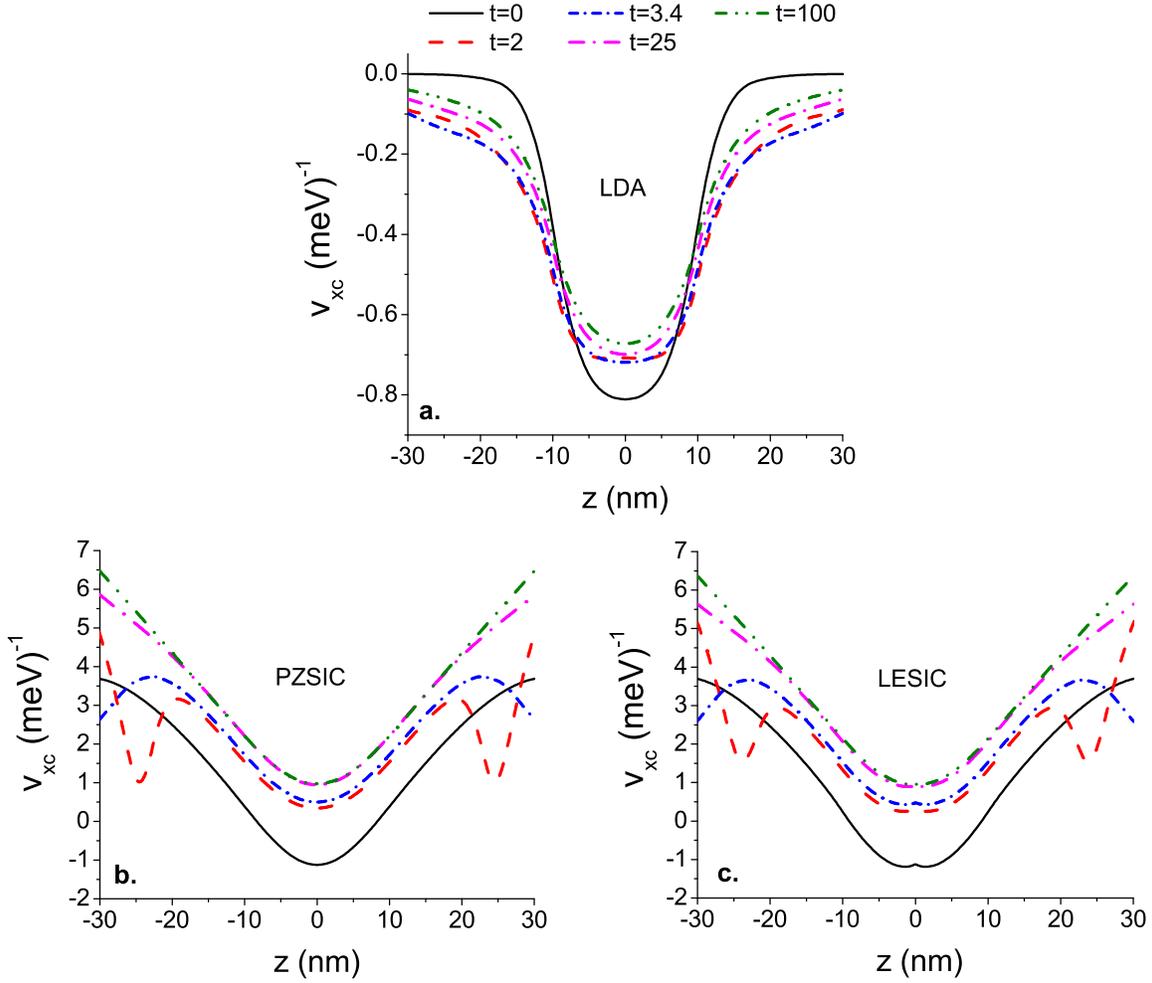}
\caption{[Color online] Snapshots of the time-dependent LDA, PZSIC and LESIC
XC potentials for the electron escape process shown in Figure \ref{fig2}.
The ionization of the second subband produces a jump of the XC SIC potentials
which builds up via a step structure that moves rapidly away from the well.
\label{fig3}}
\end{center}
\end{figure}

Figures \ref{fig3}a, \ref{fig3}b and \ref{fig3}c show snapshots of the XC
potentials at different times, in the LDA, PZSIC and LESIC
approaches. Starting with two occupied subbands at $t=0$, the second subband
is almost completely depleted at $t=100$ and accordingly
the XC SIC potentials exhibit a rigid shift between these two situations.
As ionization sets in, step structures form and
quickly migrate away from the quantum well, along with the escaping
electrons. Similar effects were observed previously
in the EXX functional for the one-dimensional lithium atom \cite{kummel}.
As expected, the LDA XC potential exhibits neither the formation of the steps nor the resulting jump.

Figure \ref{fig3} also clearly illustrates the long-range behavior of the LDA and SIC potentials.
The LDA exhibits the well-known rapid exponential decay; both SIC potentials have
a much longer spatial range, approaching a $-1/z$ behavior in the barrier region, sufficiently far away from the well.

Once the appearance of steps and jumps in the time-dependent XC
potential has been confirmed, the next question is whether and how
this affects any physical
observables. Previous studies \cite{kummel} were focused mainly on the
XC potential itself. In the next section, we present an example where the
impact of the discontinuities of the time-dependent XC potential plays an
essential role, namely the dissociation process of a double quantum well.

\subsection{Dissociation of a double well}

A better understanding and more accurate description of dissociation
processes is a central issue of much modern research in DFT and TDDFT
 \cite{neepa}. Traditionally this problem has been approached
from a static point of view \cite{perdew}. Molecular dissociation displays a basic
behavior: the separated atomic systems must have integral charge,
and this must be reproduced by any
accurate XC functional, either in the static or time-dependent
case. Unfortunately, most local and semilocal functionals fail to satisfy
this requirement and lead to final dissociation products with fractional charges.

It has been shown \cite{perdew,neepa} that, as a molecule
dissociates, the XC potential develops a sharp peak at the
bond midpoint followed by the buildup of step structures.
However, these insights were obtained in a quasistatic picture,
using ground-state calculations in some model systems. Here, we use a double
quantum well system as a simple {\em time-dependent} model of a dissociating
heteroatomic molecule. We analyze both the XC
potential itself and the effect of jumps or steps on the total
number of particles placed in each side of a double well system.

We start at $t=0$ with an asymmetric double quantum well divided by
an initially very thin (0.2 nm) barrier, such that electrons can be viewed as
sharing both wells, analogous to a molecular orbital. The left well has width
32 nm, and the right one has width 8 nm, and both have the same depth of 220
meV (see lower panel of Fig. \ref{fig1}).
The system is populated with a given total number of electrons, $N_{\rm total} = N_{\rm left} + N_{\rm right}$.

For $t>0$, we let the system dissociate, separating the quantum wells from each other
by gradually increasing the width of the dividing barrier up to a width of $10$nm.
The whole process is allowed to take a total time of 250 a.u., as indicated in the lower panel of Figure \ref{fig1}.
We chose such a slow dissociation speed in order to avoid the strong density fluctuations
that would be induced if the wells were torn apart too rapidly; nevertheless, the process is far from
being adiabatic, and some charge-density oscillations cannot be avoided.
At the final separation of 10 nm at $t=250$ a.u., however,  these charge-density oscillations
have become small ripples, as we will illustrate below. For all practical purposes, the dissociation can then be considered complete.

\begin{figure}
\begin{center}
\includegraphics{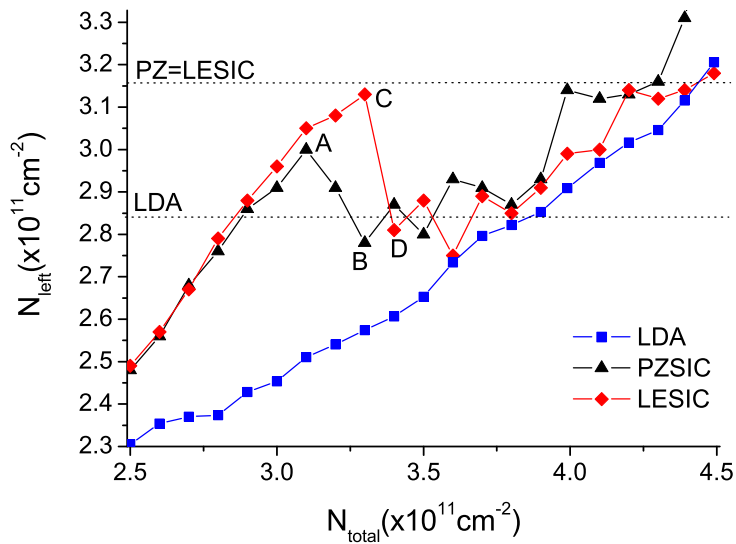}
\caption{[Color online] Number of electrons in the left well, $N_{\rm left}$,
as a function of total electron number $N_{\rm total}$
of a dissociated double quantum well. Both SIC approaches predict an abrupt decrease of $N_{\rm left}$ approaching
the region where the second subband would be filled in the isolated well (dotted lines). \label{fig4}}
\end{center}
\end{figure}

We monitor the total number of particles in the left and right quantum well,
$N_{\rm left}$ and $N_{\rm right}$, while the system dissociates.
In Figure \ref{fig4} we display $N_{\rm left}$ as a function of $N_{\rm total}$ when the separation of the
two wells is complete at $t=250$ a.u.
The horizontal dotted lines indicate the number of electrons $N_{\rm left}$ at which, in a
ground-state calculation, the second subband in the isolated left quantum well would
start to become occupied. This value of $N_{\rm left}$ is the
same for both SIC calculations, but a smaller value is predicted
in LDA. The reason for this difference is that SIC (as well as EXX) leads to larger intersubband level spacings
due to the different asymptotic behavior of the potential
\cite{proetto2} (see also Figure \ref{fig3}).

As Figure \ref{fig4} shows, the LDA predicts a continuous and rather smooth
increase of $N_{\rm left}$.
By contrast, the SIC results behave dramatically differently when $N_{\rm left}$ approaches the region where
the second subband would start to become occupied in the isolated left well. The electrons seem to resist
filling the left well, and one even sees a marked decrease of $N_{\rm left}$. If $N_{\rm total}$ continues to
increase, $N_{\rm left}$ picks up again, and eventually crosses the threshold of the second subband.

\begin{figure}
  \begin{center}
   \includegraphics{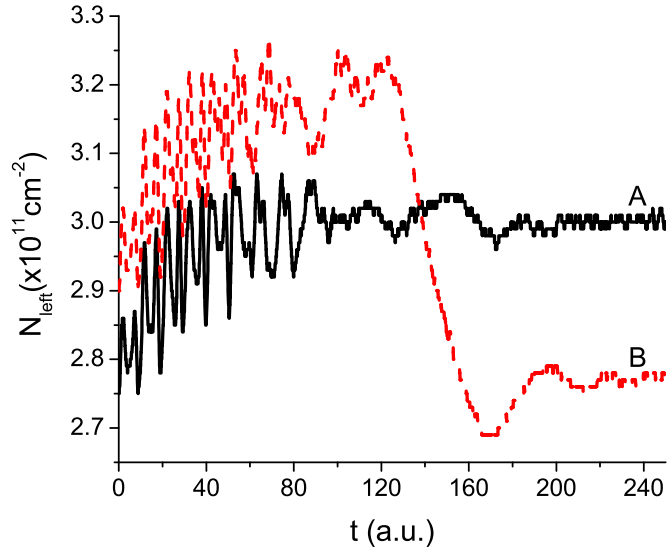}
   \caption{[Color online] Time evolution of $N_{\rm left}$ for the parameters
of point A and B indicated in Figure \ref{fig4}. The drop of particle
number in the left well occurs as soon as the system is sufficiently separated, so that the subband structure
of the isolated left well starts to become relevant.
\label{fig5}}
  \end{center}
\end{figure}

In order to understand this behavior, we plot in Figure \ref{fig5} the time evolution of $N_{\rm left}$ calculated with PZSIC,
up until the dissociated limit of $10$nm, with parameters corresponding to the points A and
B in Figure \ref{fig4}. Situation B starts with a slightly larger density in the left well,
but displays a dramatic drop as soon as the two wells are sufficiently separated. The
same behavior is found in the LESIC approach, comparing points C and D in Figure \ref{fig4}.
This indicates a repulsive force causing electrons to flow from the left
to the right well -- a clear signature of steps in the potential.

\begin{figure}
  \begin{center}
   \includegraphics{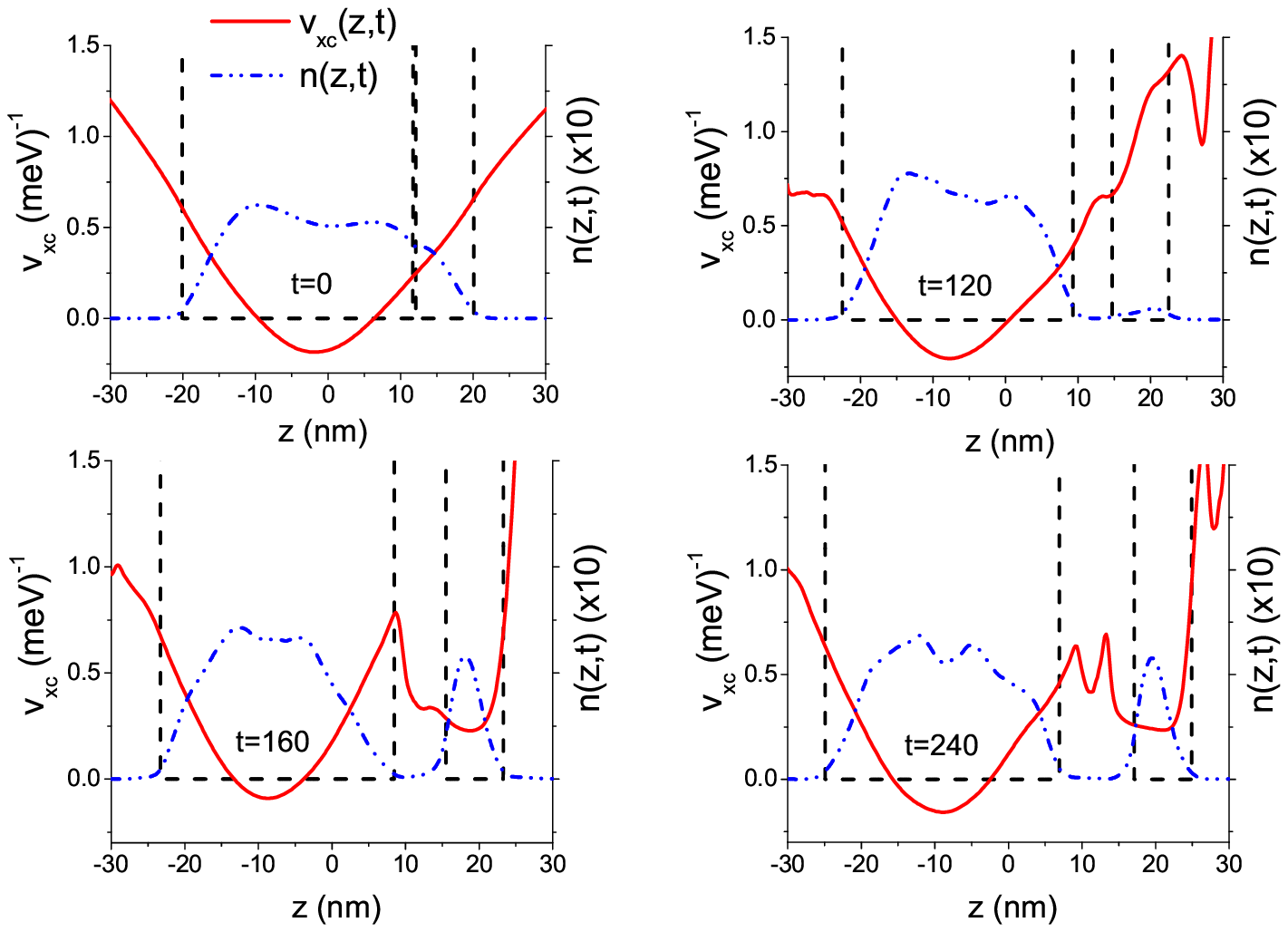}
   \caption{[Color online] Snapshots of the PZSIC XC potential corresponding
to curve B in Figure \ref{fig5}. Due to the rise of the XC potential in the
the left well relative to the right, electrons are prevented from filling
the second subband and are forced to move to the right well instead. A potential peak builds up in region between
the wells, leading to a repulsive force. The dashed lines indicate $v_{\rm ext}(z,t)$, and the dash-dotted lines show
the density $n(z,t)$. \label{fig6}}
  \end{center}
\end{figure}

In Figure \ref{fig6} we plot snapshots of the PZSIC XC potential and of the density $n(z,t)$ at different times, for parameters corresponding to
point B in Figures \ref{fig4} and \ref{fig5}, i.e., $N_{\rm total}=3.3 \times 10^{11} \rm cm^{-2}$ at the initial time.
The snapshots of the XC potential clearly illustrate the mechanism preventing the left quantum well from being
more filled: Once the double well starts to dissociate, the XC potential builds up
steps  structures, with a very pronounced sharp peak in the
region between the wells. The system resists putting
electrons in the second subband: the moment this happens, the
potential on the left side shoots up, which means that electrons flow
back to the right, as seen from the plots of the density. This effect is clearly absent in the LDA.

It is evident that in a TDDFT computational treatment of molecular
dissociation such a behavior of XC potential and density must
play a crucial role. The key physical requirement is the neutrality of
the isolated atomic system. By building up steps and jumps in the
potential, the system avoids incorrect fractional-charge
dissociation, forcing electrons to flow from an atom to another.
This is the main mechanism which restores the nature of
neutral atoms during the dissociation processes, with the correct
integral final charges.

\section{Conclusion}

The results presented here demonstrate that the essential features of the
LDA plus SIC approach, namely correct asymptotic behavior and discontinuities
upon change of particle number, appear to carry over from static DFT into the time domain.
Comparing two different formulations of SIC, PZ and LE, we find only very
little difference between them. Our TDDFT calculations were performed in the
TDKLI approximation, which leads to an adiabatic XC potential that is
state independent. It is known
that the TDKLI approach may lead to difficulties with violations of the zero-force theorem
\cite{mundt}, but in our calculations this was not a problem (see also Ref. \cite{wijewardane}).

The time-dependent SIC was used in simulating the electron dynamics in doped
semiconductor quantum well systems. We first studied electron escape of a single
square well upon sudden reduction of the well depth. The main effect of SIC is
here due to its asymptotic behavior, which in general makes electron escape proceed
at a slower pace compared to the LDA. We observed clear indications of a buildup of
a step structure in the XC potential, eventually leading to a discontinuity.
These observations are in agreement with what K\"ummel and coworkers found earlier
for their model atoms \cite{lein,kummel}.

The most dramatic effect of the XC potential steps and discontinuities was observed
in the simulated dissociation of an asymmetric double quantum well. This system can
be compared with a dissociating heteroatomic molecule. Of course, quantum wells are extended
systems, so that it is not possible to observe the absolute change in particle number.
The corresponding feature in quantum wells is instead the subband population. We found that
the dissociation of the double well depends dramatically on the electronic structure of the
resulting isolated single wells, namely, whether the second subband in the wider well would
be populated or not. It turns out that in SIC the system tends to resist populating the
second subband level as long as possible; a pronounced peak structure develops between the
two wells, and the depth of the XC potential of the wider well jumps with respect to the
potential in the other well. As a result, the total particle numbers of the left and the
right well after dissociation are dramatically different from what one would find with
the LDA.

Our calculations were carried out for relatively slowly dissociating quantum wells. Nevertheless,
we were quite far away from the adiabatic limit, as indicated by the presence of noticeable
fluctuations and charge-density oscillations which, however, subsided once the wells were
properly separated. The step structures and jumps of the XC potential become more and more
difficult to discern the more nonadiabatic and abrupt the dynamics becomes, and tend to be
washed out by strong fluctuations of densities and currents.

Further systematic studies to identify
dynamical regimes and observables for which the time-dependent XC potential discontinuities
are relevant remain an important task. However, there is no doubt that the effect is
crucial for generic dissociation or fragmentation processes, and can be captured in TDDFT by
simple, self-interaction corrected adiabatic functionals.

{\bf Acknowledgment.}
CAU was supported by NSF Grant No. DMR-0553485 and by Research Corporation.
DV and KC were supported by FAPESP and CNPq.




\clearpage

\clearpage




\clearpage


\end{document}